\documentclass[journal,twoside,web]{ieeecolor}
\usepackage{generic}
\usepackage{cite}
\usepackage[dvipsnames]{xcolor}
\usepackage{amsmath,amssymb,amsfonts}
\usepackage{algorithmic}
\usepackage{bbm}
\usepackage{graphicx}
\usepackage[outdir=./]{epstopdf}
\usepackage{epsf}
\usepackage{pict2e} 
\usepackage{overpic}
\usepackage{tikz}
\usepackage[outline]{contour}
\usepackage{float}
\usetikzlibrary{shapes.geometric, arrows, positioning, calc, matrix,fit}
\usepackage{textcomp}

\graphicspath{{figures/}}

\begin{document}
\title{Machine Learning and Feature Engineering for Predicting Pulse Status during Chest Compressions}
\author{Diya Sashidhar$^{*,\dag}$, 
Heemun Kwok$^{**,\dag}$, Jason Coult$^{\dag}$, Jen Blackwood$^{\dag}$, Peter Kudenchuck$^{\ddag,\dag}$, Shiv Bhandari$^{\dag}$, Thomas Rea$^{\S,\dag}$, and J. Nathan Kutz$^{*,\dag}$ 
\thanks{This work was supported in part by the American Heart Association under Grant 19SFRN34930005.}
\thanks{$^{*}$ Department of Applied Mathematics, University of Washington, Seattle, WA.  98195.   (email: {dsashid@uw.edu})} 
\thanks{$^\dag$ Center for Progress in Resuscitation, University of Washington, Seattle, WA.  98195. }
\thanks{$^{**}$ Department of Emergency Medicine, University of Washington, Seattle, WA.  98195.  } 
\thanks{$^\ddag$ Heart Institute, University of Washington, Seattle, WA.  98195. }
\thanks{$^\S$ Harborview Medical Center, and General Internal Medicine, University of Washington, Seattle, WA. 98195. }}

\maketitle

\begin{abstract}
\\
\textit{Objective:} Current resuscitation protocols require pausing chest compressions during cardiopulmonary resuscitation (CPR) to check for a pulse. However, pausing CPR during a pulseless rhythm can worsen patient outcome.  Our objective is to design an ECG-based algorithm that predicts pulse status during uninterrupted CPR and evaluated its performance. 
\textit{Methods:} We evaluated 383 patients being treated for out-of-hospital cardiac arrest using defibrillators with real-time ECG, CPR monitoring, and audio recordings. We collected paired immediately adjacent ECG segments having an organized rhythm.   Segments were collected during the 10-second period of ongoing CPR just prior to a pulse check, and 5-second segments without CPR during the pulse check. ECG segments with or without a pulse were identified by the audio annotation of a paramedic’s pulse check findings and recorded blood pressures. Patients were randomly divided into 60\% training and 40\% test groups. From training data, we developed an algorithm to predict the clinical pulse status based on the wavelet transform of the bandpass-filtered ECG, applying principle component analysis.   We then trained a linear discriminant model using 3 principle component modes as input features. Model performance was evaluated on test group segments with and without CPR using receiver operating curves overall and according to the initial arrest rhythm.
\textit{Results:} There were 230 patients (540 pulse checks) in the training set and 153 patients (372 pulse checks) in the test set. Overall 38\% (351/912) of checks had a spontaneous pulse.  The areas under the receiver operating characteristic curve (AUCs) for predicting pulse status with and without CPR on test data were 0.84 and 0.89, respectively.  
\textit{Conclusion:} A novel ECG-based algorithm demonstrates potential to improve resuscitation by predicting presence of a spontaneous pulse without pausing CPR.   
\textit{Significance:}   Our  ECG-based algorithm predicts pulse status during uninterrupted CPR, allowing for CPR to proceed unimpeded by pauses to check for a pulse and potentially improving resuscitation performance. 
\end{abstract}


\section{Introduction}

Machine learning (ML) and artificial intelligence (AI) algorithms are transforming the scientific landscape~\cite{goodfellow2016deep,brunton2019data}.  From self-driving cars and autonomous vehicles to digital twins and manufacturing, there are few scientific and engineering disciplines that have not been profoundly impacted by the rise of machine learning and AI methods.   Medicine is no exception, with a significant growth of machine learning and AI methods developed for applications ranging from imaging \cite{wernick2010machine,erickson2017machine} to personalized medicine \cite{holzinger2014trends}.   We specifically develop a novel ECG-based ML algorithm that demonstrates the potential to improve resuscitation by predicting presence of a spontaneous pulse without pausing CPR.

\begin{figure}[b]
\begin{tikzpicture}
    \node[anchor=south west,inner sep=0] at (0,0){\includegraphics[width = .5\textwidth]{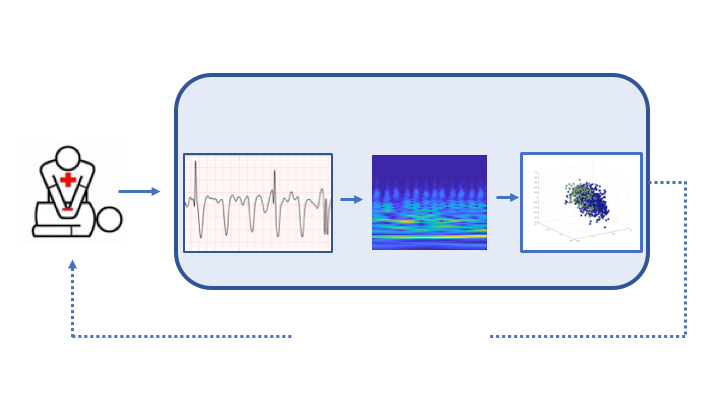}};
    \node at (.9,3.5){CPR};
    \node at (3.25,2.55) {\includegraphics[width = .107\textwidth]{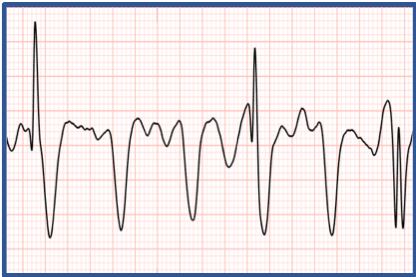}};
    \node at (3.25,3.5) {ECG};
    \node at (5.4,3.85) {Wavelet};
    \node at (5.4,3.5) {Transform};
    \node at (7.35,2.55) {\includegraphics[width = .07\textwidth]{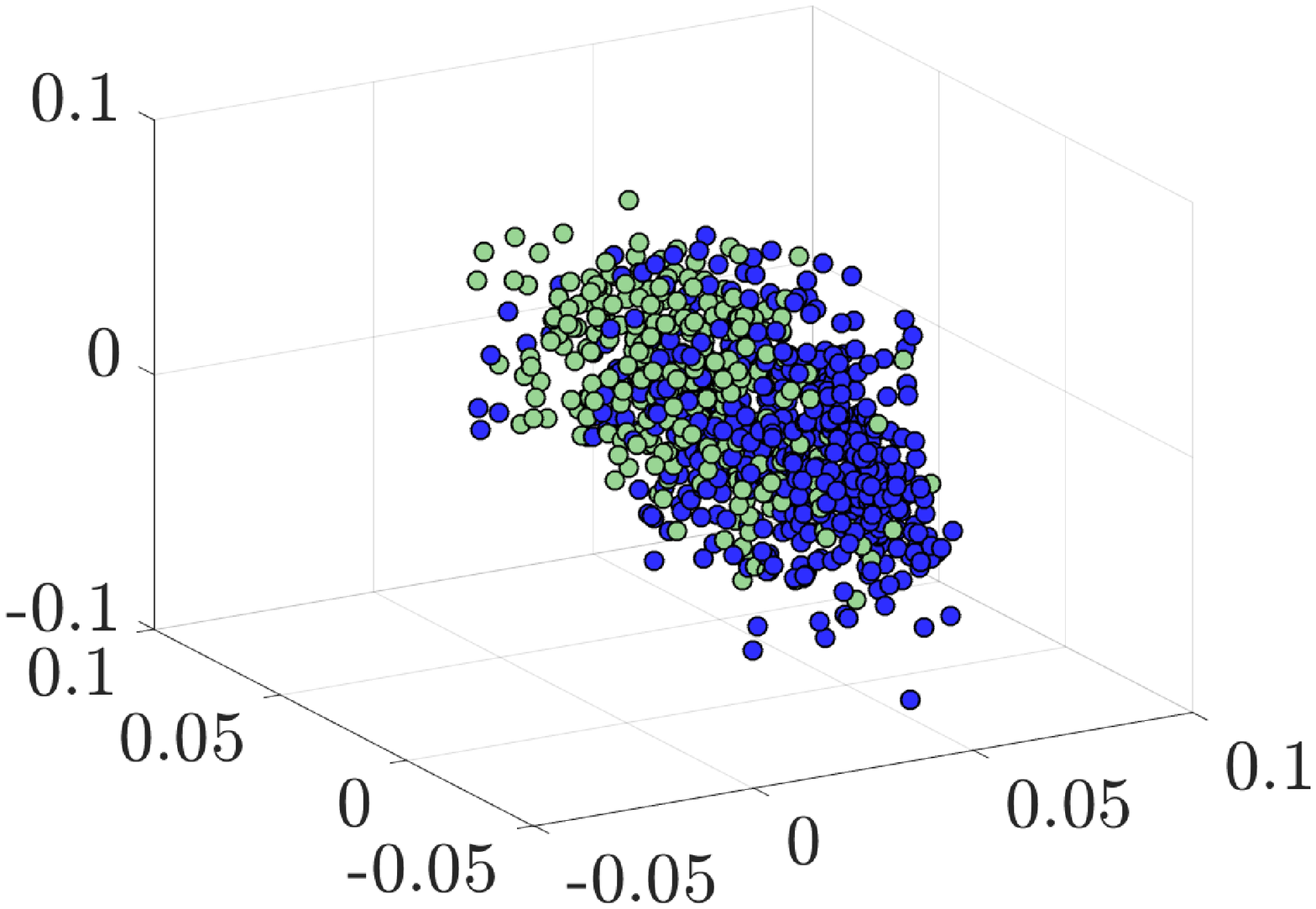}};
    \node at (7.35,3.5){PCA};
    \node at (6.95,2.75){\scriptsize {\color{ForestGreen}Pulse}};
    \node at (7.6 ,2.3){\scriptsize {\color{blue}No Pulse}};
    \node at (4.9,.9){Pulse Status};
    \end{tikzpicture}
    \caption{Overview of algorithmic architecture for real-time pulse status classification.  Time-series from ECG data is transformed to a feature space generated from wavelet analysis.  Dominant correlated features in this space are used for assessment of pulse status, with accuracies approaching nearly 90\%.}
    \label{fig:overview}
\end{figure}
\begin{table*}[t]
\centering
 \begin{tabular}{l | c  c c c} 
 Patient Characteristics& Training (N=230) & Test (N=153)& All (N=383)& Significance\\
  \hline
  Initial Rhythm &&&& \\
  \hspace{2mm} Ventricular Fibrillation (VF) &192(83.5)&128(83.7)&320(83.6)& 0.96\\
  \hspace{2mm} Asystole &11(4.7)&5 (2.2)& 16(4.2)& 0.47\\
  \hspace{2mm} Pulseless Electrical Activity (PEA) &26(11.3)&20(13.1)&46(12.0)& 0.6\\
  \hspace{2mm} Indeterminate &1(0.4)&0(0.0)&1(0.2)& *\\
  Female, n (\%) &63(27.4) & 42(27.5)  &105(27.4) &0.72\\ 
  Age, median (IQR) & 64(53, 74.5)  & 62(50.5, 73)  &63(52, 74) & 0.62\\
  Cardiac etiology, n(\%)&  201(87.4) &128(83.7) &329(85.9) &0.37\\

  Location, n(\%)  &&&& \\
 \hspace{2mm} Home &155(67.4)& 99(64.7)&  254(66.3)& 0.53\\
 \hspace{2mm}Public& 61(26.5) & 43(28.1) & 104(27.2) &0.84\\
  \hspace{2mm}Nursing Home &13(5.7) &12(7.8) &25(6.5) &0.4\\

  Arrest before EMS arrival, n(\%)& 215(92.6) &141(92.2) 356(93)& 0.73\\  
  Witnessed, n(\%)& 164(71.3) &100(65.4) &264(68.9) &0.19\\
  Bystander CPR, n(\%)& 151(65.7) &97(63.4)  &248(64.8) &0.52\\
    EMS Response (minutes), median (IQR)  &5.1(4.3, 6.9) &5.2(4.4, 6.8) &5.2(4.4, 6.9)& 0.47\\
  Total shocks, median (IQR)&  3(1, 7) &2(1, 3.5) &2(1, 5)& 0.06\\
    ROSC, n(\%)&  156(67.8) &109(71.2) &265(69.2) &0.58\\
      Admit to hospital, n(\%)& 164(71.3) &110(71.9)& 274(71.5) &0.42\\

    Survive to hospital discharge, n(\%)& 108(47) &69(45.1) & 177(46.2) &0.51\\
\end{tabular}
\label{table:demo}
\caption{Patient Demographics. Note: Significance was calculated using Mann-Whitney U test for continuous variables and chi-square for proportions}
\end{table*}

ECG data and AI/ML algorithms are ideally poised to better inform resuscitation efforts in patients with cardiac arrest. Hundreds of thousands of people suffer an out-of-hospital cardiac arrest (OHCA) each year \cite{benjamin2018heart}. Patients who suffer a cardiac arrest require time-critical, life-saving interventions for successful resuscitation.  These interventions include cardiopulmonary resuscitation (CPR), which is a combination of chest compressions, ventilations, and advanced airway management. Additional interventions, such as electrical defibrillation and drug administration, depend upon knowledge of the cardiac rhythm and the presence of a pulse beneath CPR.  Ideally, the rhythm and presence of a pulse would be monitored continuously during resuscitation to help guide treatment decisions.  However, chest compressions preclude the accurate assessment of rhythm or a \textit{spontaneous} pulse, because compressions cause artifact in the electrocardiogram (ECG) and can generate some measure of pulse and blood pressure themselves.  Conversely, pausing CPR to assess these parameters comes at the price of depriving the patient, if pulseless, of needed hemodynamic support and can adversely affect outcome.  Thus current guidelines reflect a balance between interrupting CPR to manually assess a patient’s pulse status versus performing continuous chest compressions in recommending an interruption in chest compressions for rhythm and pulse assessment only once every 2 minutes \cite{kleinman2015part,link2015part}.

Although these pauses are intended to be brief in duration, interruptions in CPR to evaluate rhythm and pulse may take up to 20 seconds or longer, with increasing duration inversely associated with chances of survival \cite{yu2002adverse,cheskes2011perishock}.  The ability to detect a spontaneous pulse in real time during ongoing CPR could better inform care providers of the patient’s clinical status throughout resuscitation and afford more time-sensitive, better-informed treatment decisions to achieve improved clinical outcomes.   

In the current study, we introduce a scalable approach that integrates automated feature selection with novel machine learning techniques (see Figure \ref{fig:overview}). This approach uses wavelet transforms followed by principal component analysis, allowing the potential capture of subtle ECG morphologies while reducing the dimensionality of the model. Using only three dimensions of the reduced wavelet transform, we evaluate and describe a diverse number of ML models, including deep learning architectures, that predict the pulse status of organized rhythm segments both with and without CPR.  We hypothesized that this novel approach would provide accurate detection of spontaneous pulse during ongoing CPR.

\section{Methods}

 \subsection{Description of Data}
This retrospective observational study evaluated a convenience sample of 383 patients who experienced OHCA and underwent attempted resuscitation by emergency medical services (EMS) in King County, WA between 2006 and 2014 and had a complete electronic dataset available from the resuscitation.  Demographic, clinical, and outcome characteristics of the study cohort are provided in Table \ref{table:demo}. The defibrillator recording from each case included continuous measurement of the ECG and thoracic impedance, and an audio channel containing EMS narration of events during the course of resuscitation.  Each defibrillator recording was reviewed by experts to continuously classify CPR \cite{kwok2016accurate,coult2019ventricular}, rhythm \cite{kerber1997automatic}, and when there was clinical evidence of return of spontaneous circulation (ROSC) \cite{kwok2020electrocardiogram}.  

\begin{figure}[t]
      \begin{overpic}
    [width=.5\textwidth]{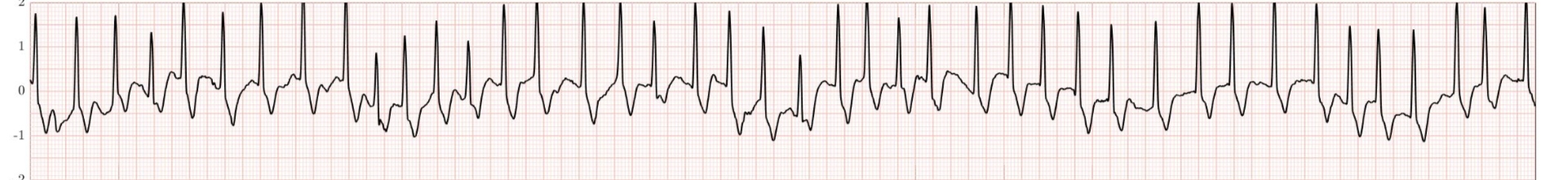}
    \put(25,13.5){\color{blue}CPR},
    \put(50,-5.5){Time(s)}
    \put(1.7,-2){\tiny 0},
    \put(7,-2){\tiny 1}
    \put(12.7,-2){\tiny 2}
    \put(18.5,-2){\tiny 3}
    \put(24,-2){\tiny 4}
    \put(29.7,-2){\tiny 5}
    \put(35.3,-2){\tiny 6}
    \put(41,-2){\tiny 7}
    \put(46.7,-2){\tiny 8}
    \put(52.4,-2){\tiny 9}
    \put(57.4,-2){\tiny 10}
    \put(63,-2){\tiny 11}
    \put(68.7,-2){\tiny 12}
    \put(74.5,-2){\tiny 13}
    \put(80,-2){\tiny 14}
    \put(85,-2){\tiny 15}
    \put(90,-2){\tiny 16}
    \put(95,-2){\tiny 17}
    \put(-3,-3.5){\rotatebox{90}{\small Voltage(mV)}}
    \put(-.5,-2.7){\color{blue}\linethickness{0.5mm}
\polygon(1,2.5)(59,2.5)(59,14)(1,14)}
    \put(80,13.5){\color{red} No CPR}
    \put(-.5,-2.7){\color{red}\linethickness{0.5mm}
\polygon(70,2.5)(97,2.5)(97,14)(70,14)}
  \vspace{1cm}
  \put(28,-6){$\Downarrow$}
  \put(82,-6){$\Downarrow$}
  \end{overpic}

  \begin{minipage}{.26\textwidth}
  \vspace{.8cm}
  \hspace{.07 cm}
    \begin{overpic}
    [width=1.09\textwidth]{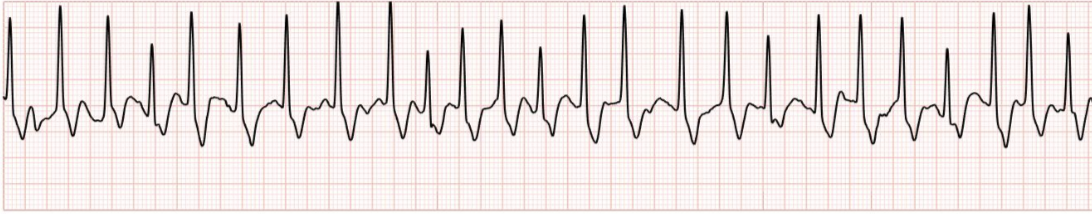}
    \end{overpic}
  \end{minipage}
  \begin{minipage}{.22\textwidth}
    \vspace{.8cm}
    \hspace{1.35cm}
    \begin{overpic}
    [width=.64\textwidth]{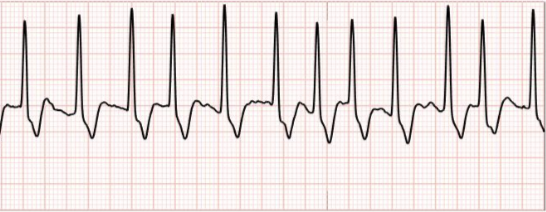}
  \end{overpic}
  \end{minipage}
  \caption{ECG after using 4-tap Butterworth filter process}
  \label{fig:filtered}
\end{figure}
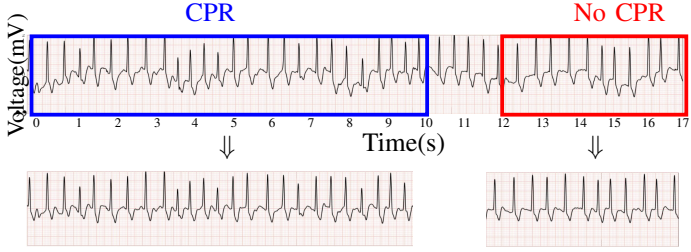

We developed and evaluated the algorithm on adjacent ECG segments surrounding the time of rhythm and pulse assessment during the immediate preceding period of ongoing CPR and when CPR was withheld for this purpose.  Specifically, for each pulse check, a 10-second ECG segment was collected during the period of ongoing CPR just prior to the pulse check, and a 5-second adjacent segment was collected without CPR during the pulse check. We assumed pulse status was unlikely to have changed during this very short interval (Figure \ref{fig:filtered}). The number of pulse checks depended upon the course of resuscitation and was therefore variable across patients.  For example, a patient who achieved ROSC quickly may have only one pulse check, while another patient who did not ultimately survive may have had many pulse checks. To prevent the overrepresentation of data from any single patient, we set a maximum of three 15-second periods from each category (\textit{Pulse} or \textit{Pulseless}) per patient.

We extracted a pair of ECG segments (with and without CPR) from a total of 912 rhythm/pulse checks.  Patients from whom these ECG segments were collected were randomized into training (60\% of patients) and test (40\% of patients) groups. In the training set, there were 230 patients with 540 segment pairs (with and without CPR), and in the test set, there were 153 patients with 372 segment pairs. Training and test patients had similar demographic and resuscitation characteristics (see Table \ref{table:demo}).  The percentage of \textit{Pulse} vs. \textit{Pulseless} segments were similar in the training (39\% \textit{Pulse}) and test (37\% \textit{Pulse}) sets.

 \subsection{Data Preprocessing and Noise Removal}
Data was collected from MRx and Forerunner 3 (Philips Healthcare, Bothell, WA), and Lifepak 12 and Lifepak 15 (Physio-Control, Redmond, WA) automated biphasic defibrillators, with ECG sampling rates ranging from 125 to 250 Hz. We homogenized all ECG data to a sampling rate of 250 Hz. We applied filtering to remove high-frequency electrical noise and low-frequency drift using a fourth-order Butterworth bandpass filter from 1-40 Hz implemented with a forwards-backwards implementation to preserve linear phase.  Of note, this filter was not designed to remove artifact from CPR, which is concentrated primarily at the frequency of chest compressions (2 Hz) but can exhibit harmonic artifact (e.g. up to 20 Hz) that can overlap with underlying heart activity \cite{gong_otherfilters,coult2019ventricular}.  Other studies have attempted to reduce CPR artifact using adaptive filters, Kalman filters, and notch filters \cite{gong_otherfilters,ARAMENDI2007115}. In contrast, rather than attempt to remove CPR artifact via filtering, the current method employs a combination of  time-frequency analysis and machine learning to allow data-driven exclusion of frequencies confounded by CPR artifacts.  \\

\subsection{Algorithm Development}

\subsubsection{Wavelet Transforms}
We used wavelets as a decomposition method for analysis and representation of time-frequency signals \cite{daubechies1992ten,mallat1999wavelet}. As the mathematical structure for constructing the multi-resolution analysis of data such as the ECG time series or spatial data such as images\cite{mallat1999wavelet,kutz2013data}, wavelets are foundational in the representation of data that have multiscale spatiotemporal features. Leading image compression technologies such as JPEG 2000 are wavelet-based encoding schemes which leverage multi-resolution analysis to sparsely represent important image features for the compression process \cite{lewis1992image}. Importantly, wavelets provide a recursive architecture which extracts dominant time-frequency features within prescribed time-bins, much like the windowed Fourier transform or Gabor transform \cite{kutz2013data}.
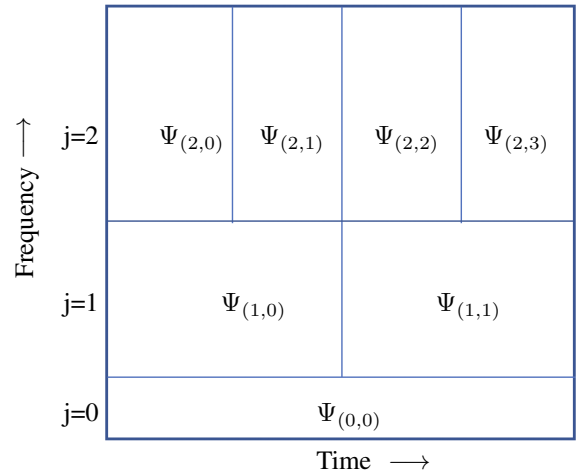
\begin{figure}[t]
\centering
\begin{overpic}
    [width = .35\textwidth]{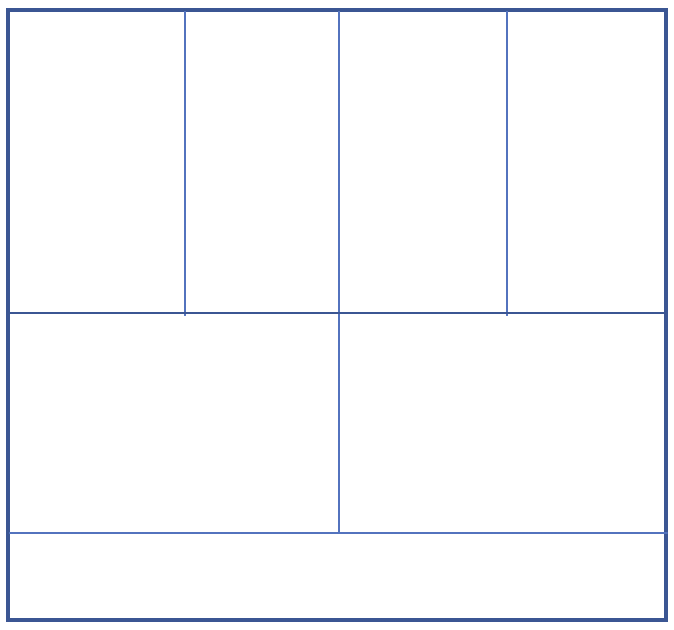}
    \put(45,5){$\Psi_{\left(0,0\right)}$}
    \put(-8,5){j=0}
    \put(25,28){$\Psi_{\left(1,0\right)}$}
    \put(70,28){$\Psi_{\left(1,1\right)}$}
    \put(-8,28){j=1}
    \put(12,63){$\Psi_{\left(2,0\right)}$}
    \put(33,63){$\Psi_{\left(2,1\right)}$}
    \put(57,63){$\Psi_{\left(2,2\right)}$}
    \put(80,63){$\Psi_{\left(2,3\right)}$}
    \put(-8,63){j=2}
    \put(45,-5){Time}
    \put(60,-5){$\longrightarrow$}
    \put(-18,35){\rotatebox{90}{Frequency}}
    \put(-17,60){\rotatebox{90}{$\longrightarrow$}}
\end{overpic}
\vspace*{.1in}
\caption{Wavelet Transform template}
\end{figure}

In the current investigation, the wavelet transform was used to analyze the time-frequency structure of the filtered ECG by constructing the ubiquitous scalogram, which is equivalent to the spectrogram created using Gabor transforms \cite{kutz2013data}. This scalogram provides a visual representation of the time and frequency content of the ECG data, and provides a critical feature space to extract meaningful portions of the signal related to the physiological status that may predict pulse.

  \begin{figure*}[t]
\begin{minipage}{.9\textwidth}
  \begin{minipage}{.5\textwidth}
\begin{overpic}
[width = 1.1\textwidth]{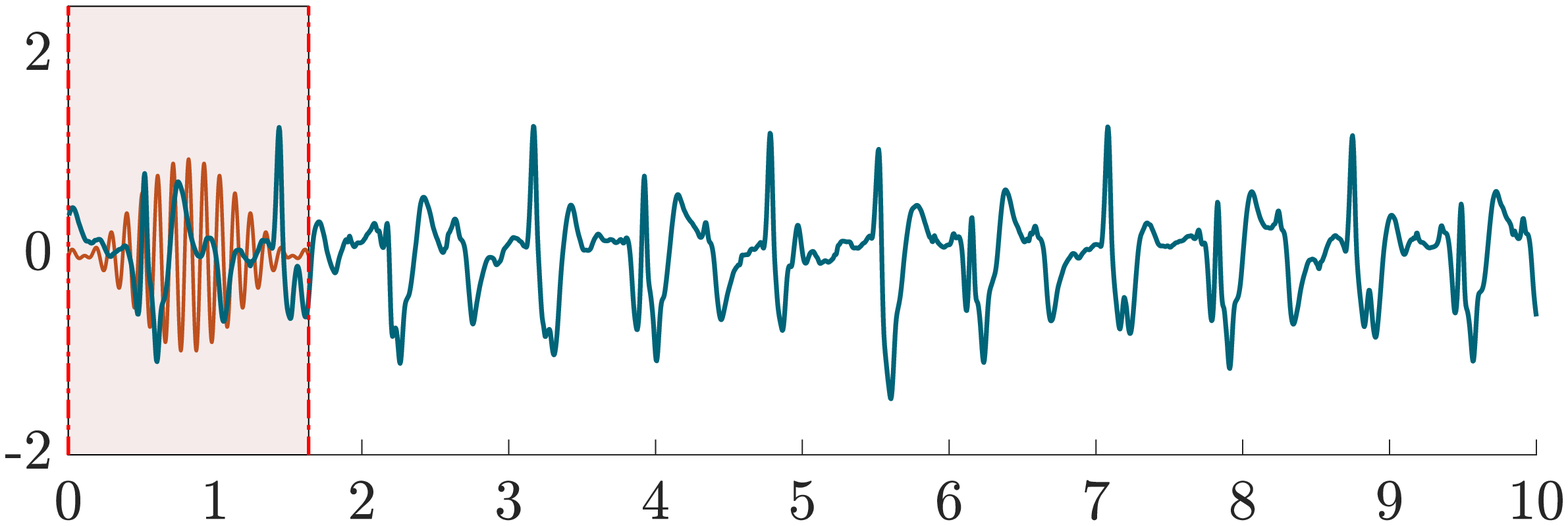} 
\put(25.2,25){\color{red}{$\longrightarrow$}}
\put(45,30){Pulse}
\put(45,-3){Time(s)}
\put(4,5){\rotatebox{90}{Voltage (mV)}}
\end{overpic}
\end{minipage}  
\hspace{1mm}  
\begin{minipage}{0.5\textwidth}
\begin{overpic}
[width=1.1\textwidth]{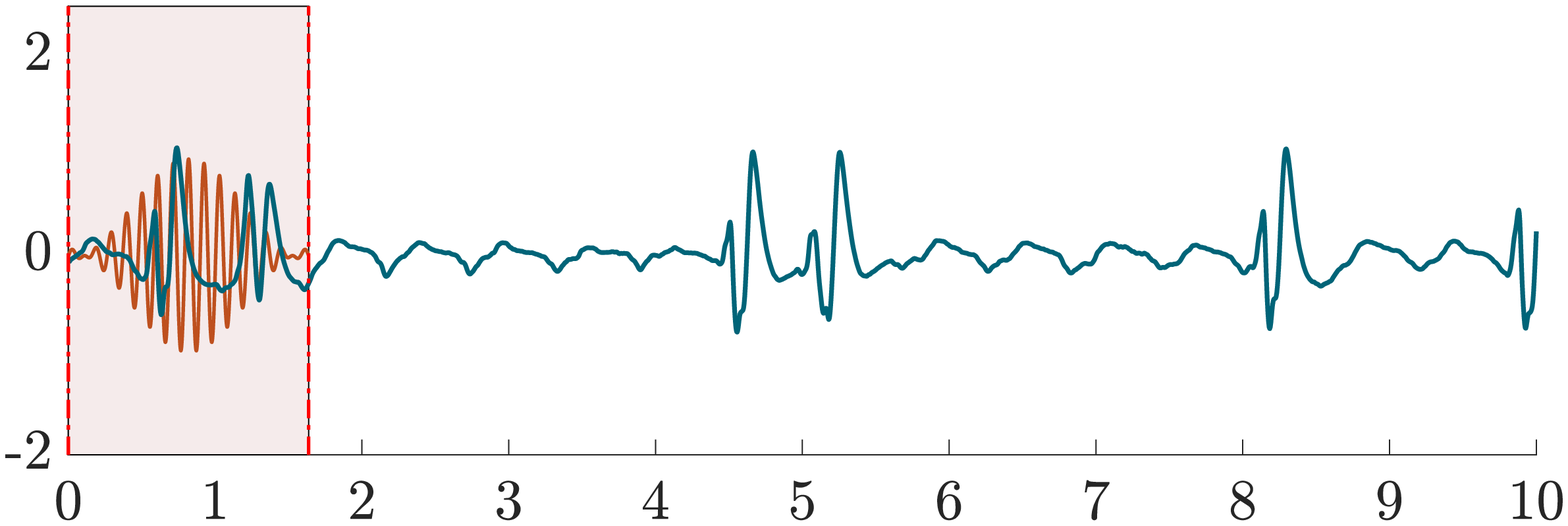}
\put(25.2,25){\color{red}{$\longrightarrow$}}
\put(45,30){No Pulse}
\put(4,5){\rotatebox{90}{Voltage (mV)}}
\put(45,-3){Time(s)}
\end{overpic}    
\end{minipage}

\vspace{1cm}

\begin{minipage}{.5\textwidth}
\centering
\hspace{2mm}  
\begin{overpic}
[width=1.1\textwidth]{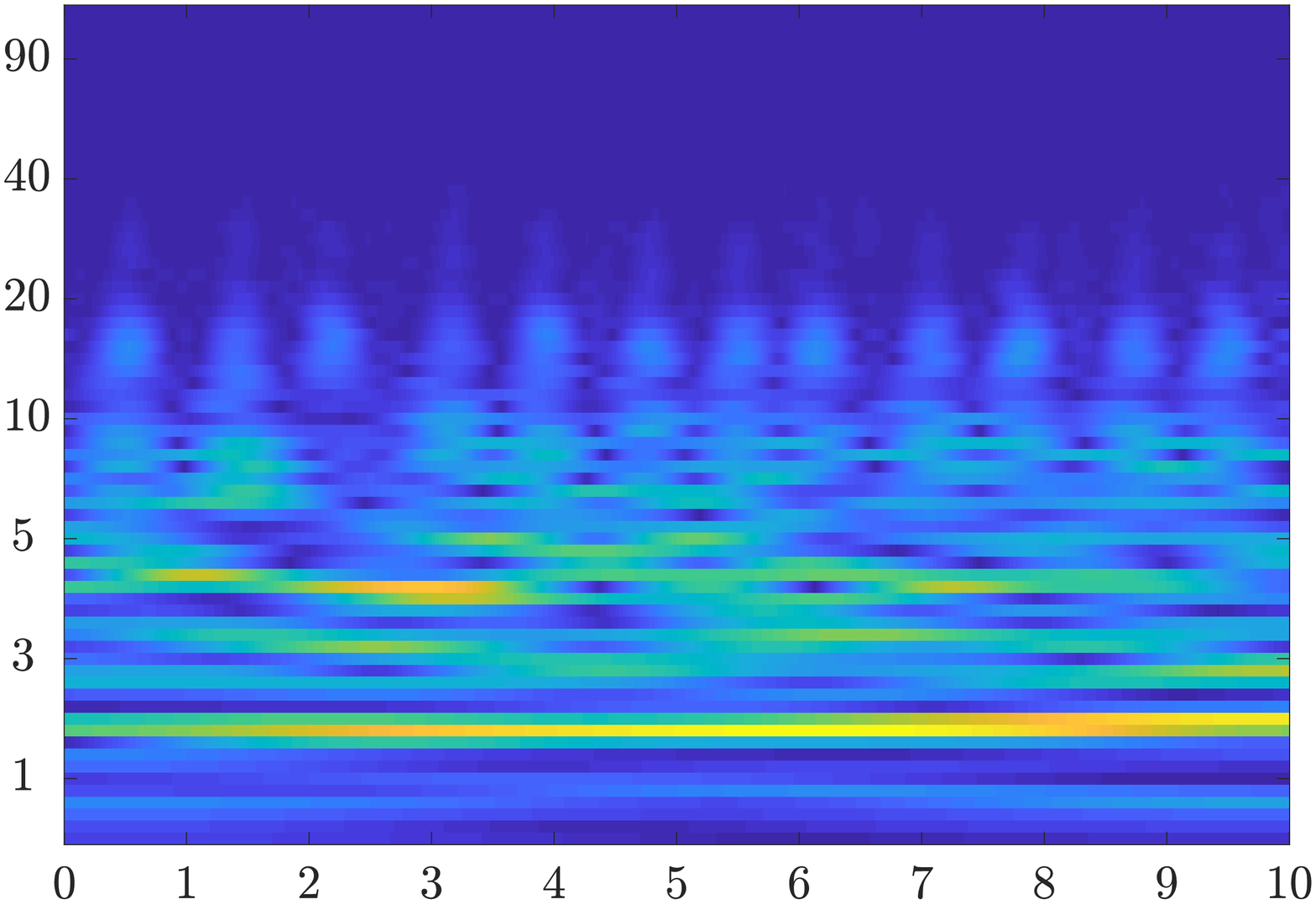}  
\put(50,65){$\Downarrow$}
\put(45,-1){Time(s)}
\put(4,25){\rotatebox{90}{Frequency (Hz)}}
\end{overpic}
\label{fig:noPulseStrip_bump}
    \end{minipage}%
    \hspace{1mm}  
    \begin{minipage}{0.5\textwidth}
    \centering
    \begin{overpic}
    [width=1.1\textwidth]{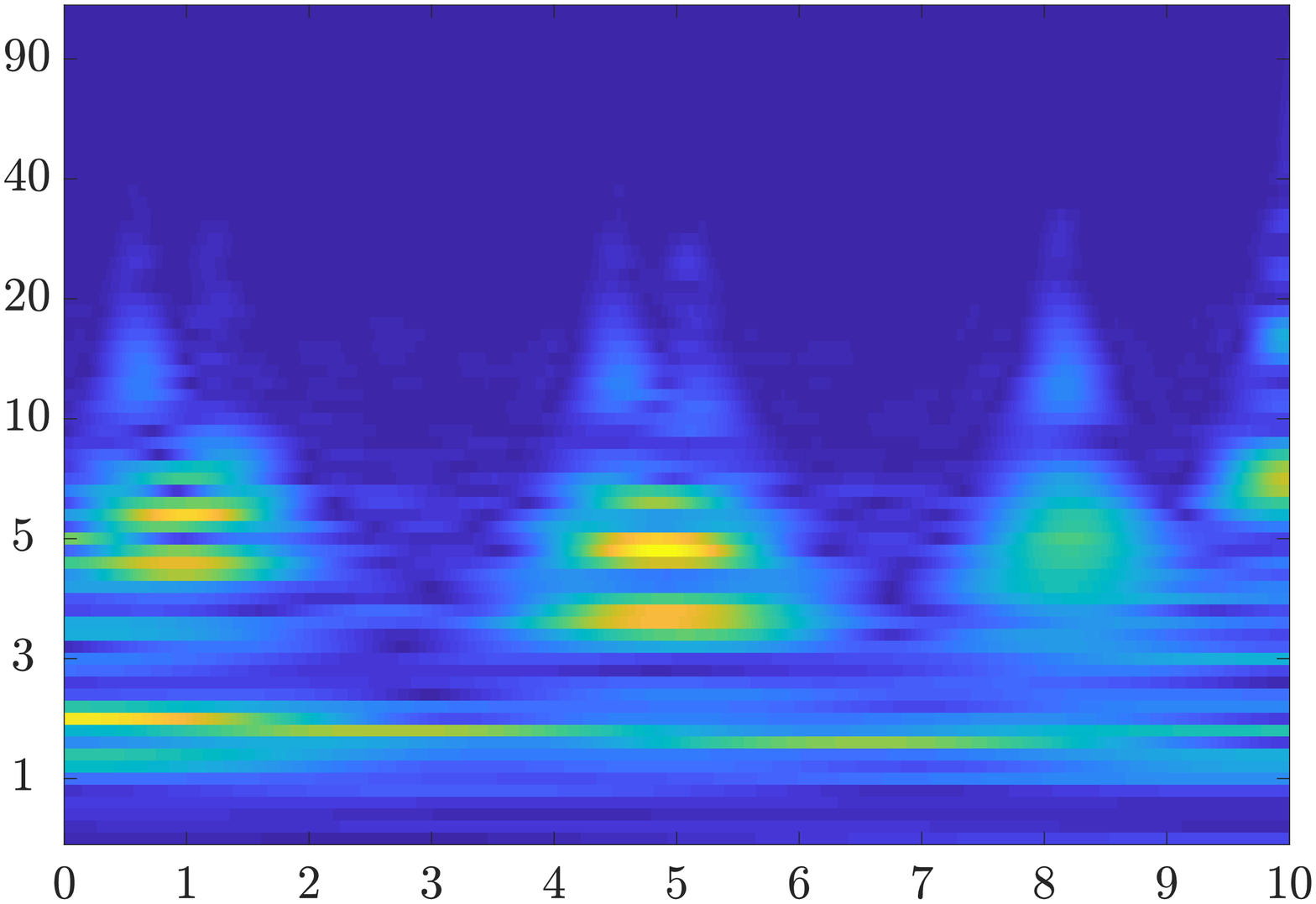}
    \put(51,65){$\Downarrow$}
    \put(45,-1){Time(s)}
    \put(4,25){\rotatebox{90}{Frequency (Hz)}}
    \end{overpic}
    \label{fig:noPulseScalo}
    \end{minipage}
\end{minipage}
\hspace{2mm}
    \begin{minipage}{0.08\textwidth}
    \includegraphics[width = 1\textwidth]{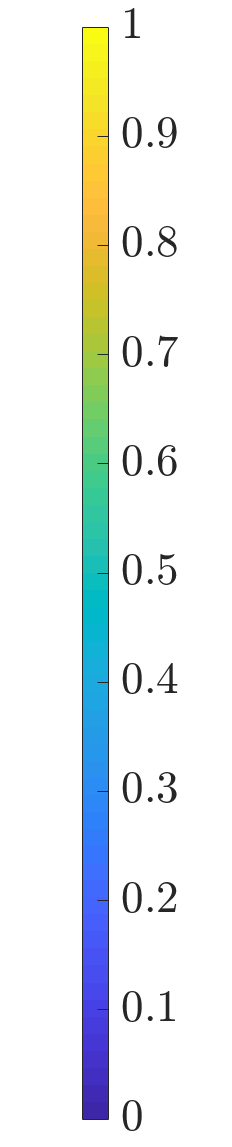}
    \end{minipage}
    \caption{Top Left: Application of bump wavelet to segment of ECG (during CPR) with ROSC label, Top Right: Application of bump wavelet to segment of ECG (during CPR) with No Pulse label, Bottom Left: Scalogram generated by convolving various scales of bump wavelet with respective ECG segment with ROSC label, Bottom Right:  Scalogram generated by convolving various scales of bump wavelet with respective ECG segment with ROSC label}
    \label{fig: scalograms}
    \end{figure*}
The wavelet transform is computed by convolving a wavelet function with the time-series recording from the ECG. The wavelet window, whose temporal width is recursively cut in half (or doubled) hierarchically, is translated across the entire ECG time series at its time-frequency information is extracted. The wavelet is represented by the functional form

\begin{align}
  \Psi_{j,k}(t) &= \frac{1}{a_{j}}\Psi\left(\frac{t-b_k}{a_{j}}\right)
  \label{eqn:DWT_psi}
\end{align}
where $a_j$ denotes the wavelet dilation parameter, $b_k$ is the center position of the wavelet window, t is the current time point, and $\Psi$ is the mother wavelet. The parameters $b_k$ is chosen so that the wavelet window translates across the entire time history of the ECG data segment. The wavelet transform convolves (1) with the ECG signal f(t) and gives a new representation that is parametrized by both $a_j$ and $b_k$:
\begin{align}
	W_\Psi(f)(a_j,b_k) = \int_{0}^{T} f(t)\bar{\Psi}_{j,k}(t)dt 
	\label{eqn:wavelet_convolve}
\end{align}
where the ECG signal is recorded for $t \in [0,T]$. The value of $W_\Psi(f)(a_j,b_k)$ for each $a_j$ and $b_k$ gives the spectral content, or energy, at a specific time-frequency location. By plotting the entire time-frequency plane, the scalogram is constructed from equation (2). Specifically, the energy of a signal (E) in a window corresponding to dilation parameter a and centered at b is given by
\begin{align}
E(a,b) = |W_\Psi(f)(a,b)|^2 .
\end{align}

To calculate the scalogram, we integrate these values across all discrete values of $a_j$ and $b_k$ used when discretizing the time-series signal.

Since wavelets can extract localized features in the time-frequency domain, we apply the wavelet transforms to the detrended ECG time-series measurements (see Figure \ref{fig: scalograms}).

Depending on the wavelet type and sifting window size, the wavelet transform can capture both the morphology of sharp peaks such as the QRS wave as well as the nuances of the P wave. The choice of the wavelet depends on the nature of the signal. Since ECGs are periodic and have oscillatory bursts as seen in QRS complexes, we choose the bump wavelet for extraction of our time-frequency features. The bump wavelet takes the Fourier form:

\begin{figure}[t]
\centering
\begin{tikzpicture}
    \node[anchor=south west,inner sep=0] at (0,0){\includegraphics[width = .5\textwidth]{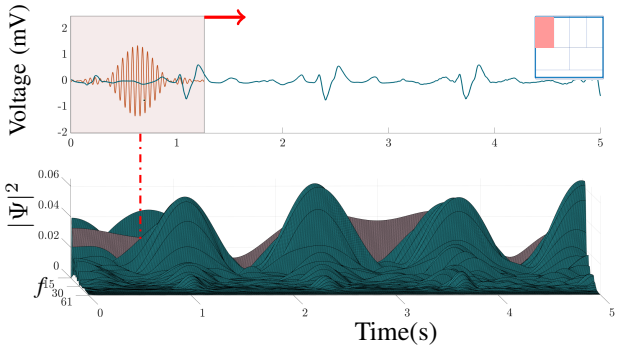}};
    \node at (7.8,3.8){\includegraphics[width = .05\textwidth]{../figures/template_DWT.png}};
    \draw[RoyalBlue](7.35,3.4) rectangle (8.25,4.2);
    \fill[red!40!white] (7.35,3.8) rectangle (7.6,4.2);
    \draw [red,thick,dash dot] (2.1,2.65) -- (2.1,1.25);
    \draw[red,->,very thick] (2.95,4.2) -- (3.5,4.2);
    \node at (5.5,0) {Time(s)};
    \node at (.5,3.5) {\rotatebox{90}{\small Voltage (mV)}};
    \node at (.5,1.7){\rotatebox{90}{$|\Psi|^2$}};
    \node at (.8,.6){\small \textit{f}};
\end{tikzpicture}
\caption{Scalogram is calculated by convolving the bump wavelet with the filtered ECG as the window translates across the time series.}
\end{figure}

\begin{align}
\hat{\Psi}(a\omega)=e^{1-\frac{1}{1-(a\omega-\mu)^2/\sigma^2}}\mathbbm{1}_{\{(\mu-\sigma)/a,(\mu+\sigma)/a\}} 
\end{align}
where $\omega$ is the current frequency, $a$ is the scale, and $\mu$ and $\sigma$ are the mean and standard deviation. The oscillatory shape of bump wavelet enables it to capture both the morphology and sharpness of the QRS wave and the dome-like structures of the P and T waves. Importantly, the wavelet transform can produce scalograms which will be used as the feature space for identifying the pulse status of the heart.

\begin{figure}[t]
\centering
   \begin{overpic}
[width=.5\textwidth]{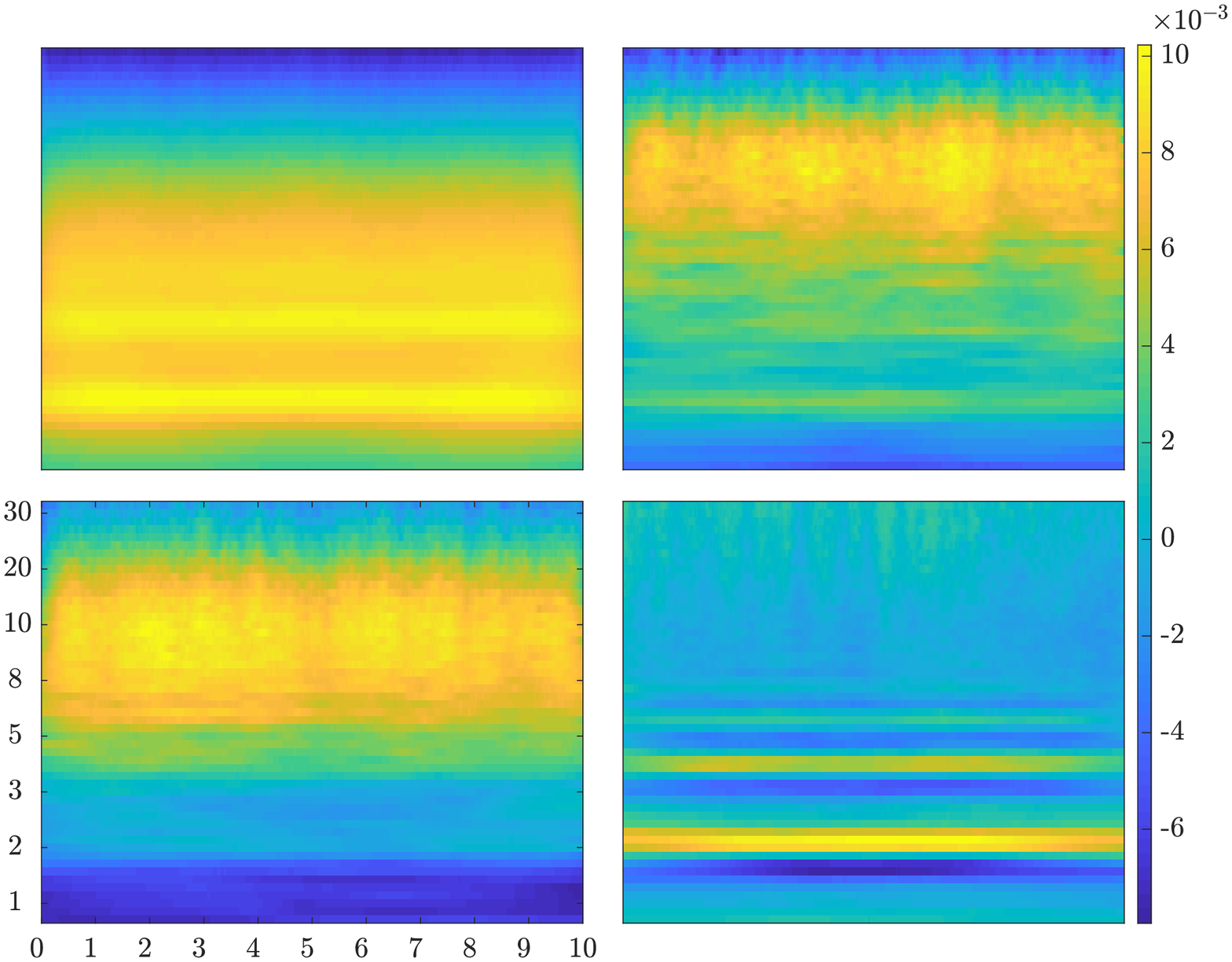}
\put(11,68){\contourlength{0.07em}\contour{black}{\color{white}\Large {Mode 1}}}
\put(54,68){\contourlength{0.07em}\contour{black}{\color{white}\Large {Mode 2}}}
\put(11,35){\contourlength{0.07em}\contour{black}{\color{white}\Large {Mode 3}}}
\put(54,35){\contourlength{0.07em}\contour{black}{\color{white}\Large {Mode 4}}}
\put(3,17){\rotatebox{90}{Frequency (Hz)}}
\put(25,2.5){Time(s)}
\put(75,13){\contour{white}{\color{blue}{\LARGE {CPR}}}}
\end{overpic}
   \centering

   \begin{overpic}
   [width=.5\textwidth, height=2in]{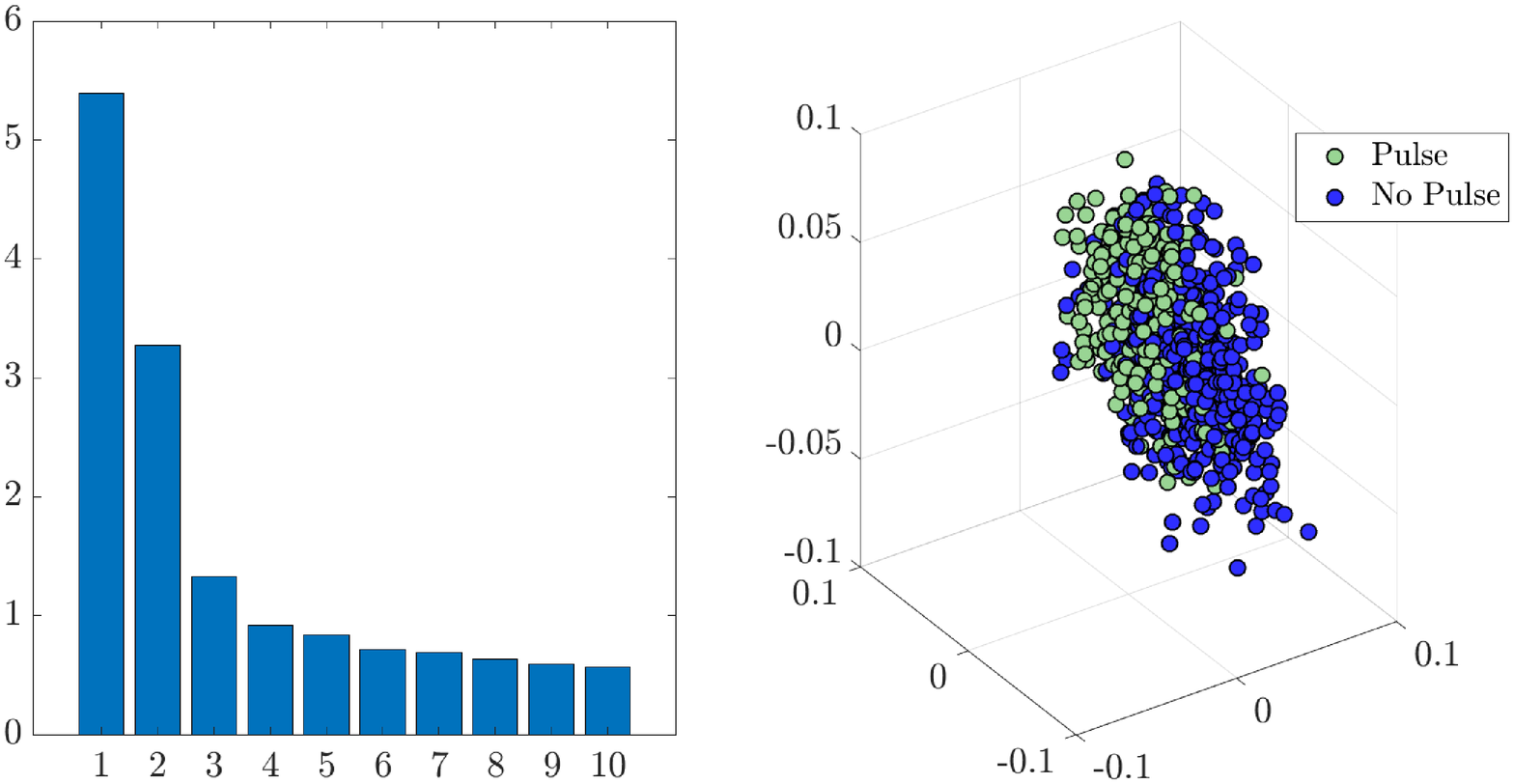}
   \put(0,25){\rotatebox{90}{\% Variance}}
   \put(20,0){Mode Number}
   
   \put(49,28){\rotatebox{90}{Mode 3}}
   
   \put(55,9){\rotatebox{-41}{Mode 2}}
   
   \put(83,1){\rotatebox{25}{Mode 1}}
     
   \end{overpic}
   \caption{Top: First four spatial modes of the system of reshaped scalograms (during CPR), Bottom Left: Normalized Singular Values of the system of reshaped scalograms, Bottom Right: Various combinations of temporal modes of system of reshaped scalograms.}
   \label{fig:singular_modes_two}
\end{figure}

\subsubsection{Feature Engineering}
After detrending and transforming ECGs into spectrograms, large amounts of artifact and voltage variability can still persist. Therefore we sought to reduce dataset dimensionality by projecting wavelet-transformed ECG data onto a low-dimensional subspace using principal component analysis (PCA). PCA is a powerful tool that captures dominant correlated structures \cite{kutz2013data}.  With just a few PCA modes, a low-dimensional representation of the system can be constructed. To remove extraneous signal in the data, we imposed a percent variance cutoff of 1\% as a criteria for selection of modes to include as features, as much of the variance of the system is captured in the first three modes. (see Figure \ref{fig:singular_modes_two}). 

\begin{figure}[t]
   \begin{overpic}
   [width=.5\textwidth]{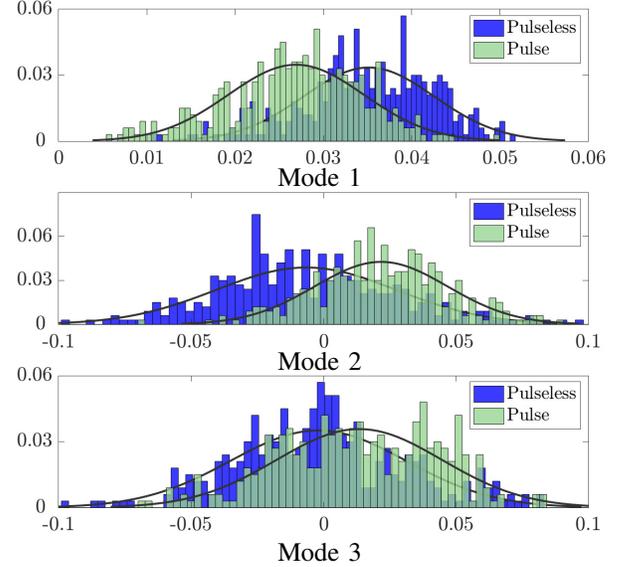}
   \put(45,2){Mode 3}
   \put(45,57){Mode 1}
   \put(45,30){Mode 2}
   \end{overpic}
   \caption{Histograms of Modes 1,2,and 3 for each label during CPR}
   \label{fig:histograms}
   \end{figure}

The distributions of modes 1-3 stratified by pulse status and CPR (see Figure \ref{fig:histograms}). provide insight into their usefulness as features. In both CPR and No CPR clips, Modes 1 and 2 have slightly-overlapping, yet distinct distributions. While the distributions in Mode 3 show more overlap, there are still slightly separated, especially with CPR clips. Thus, the addition of Mode 3, can possibly provide information that is not provided in Modes 1 and 2.

We evaluated heart rate, a variable with accepted clinical significance, as a potential feature for inclusion in the method in addition to the three modes identified through PCA.  To calculate heart rate, we first preprocessed the ECG with a higher-order (8th order) Butterworth filter and a passband of 10-40 Hz to emphasize QRS complexes. We then applied a simple peak-finding algorithm to calculate heart rate from QRS complex locations observed within the ECG segment. The heart rate (in beats/min) was then incorporated as a potential feature of the classification model.

\subsubsection{Model Selection}
\begin{figure}[t]
\centering
\begin{minipage}{0.235\textwidth}
  \begin{overpic}
        [width = .95\textwidth]{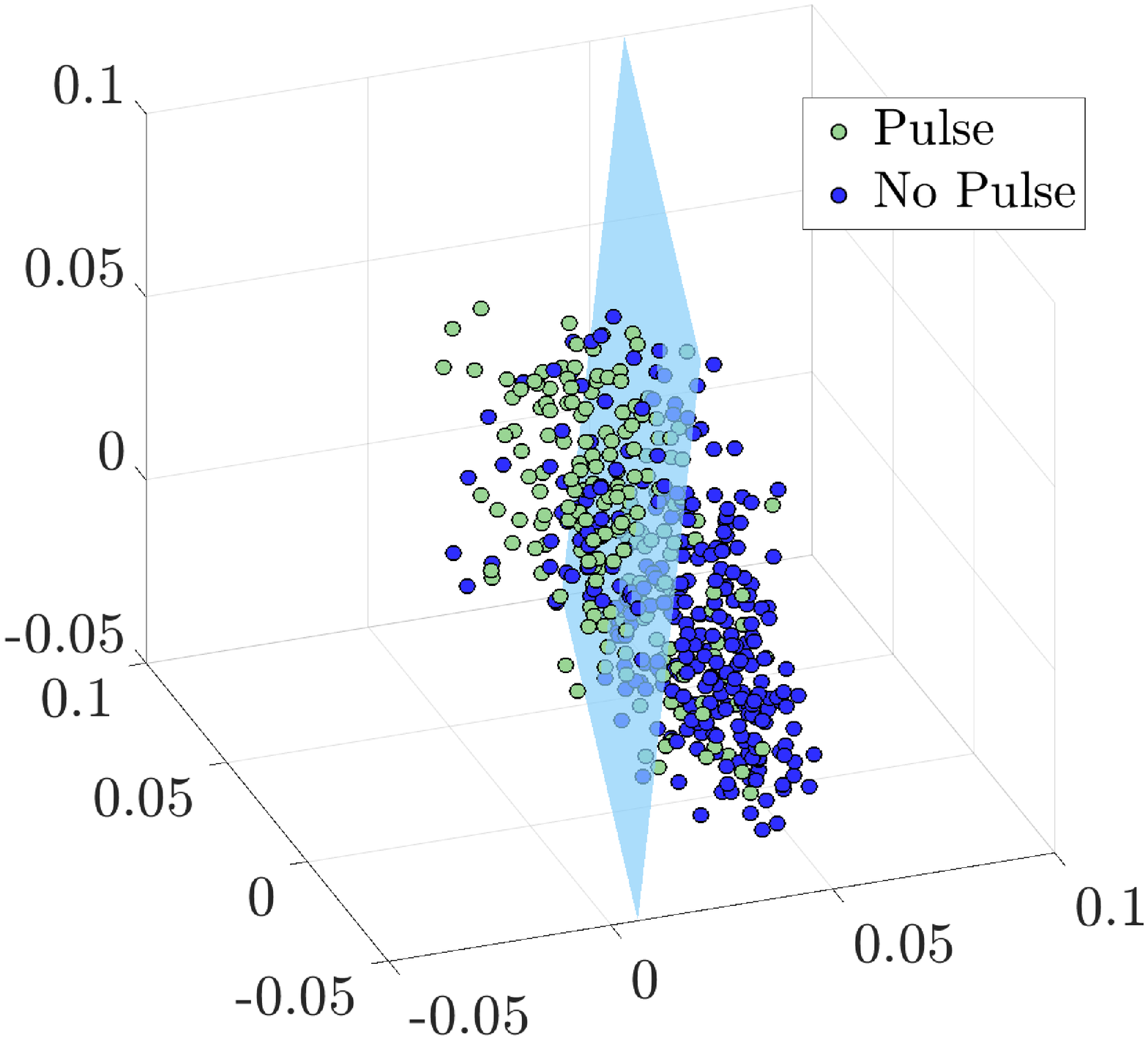}
        \put(-4,55){\rotatebox{90}{\small Mode 3}}
        \put(60,0){\rotatebox{7}{\small Mode 1}}
        \put(-4,30){\rotatebox{-55}{\small Mode 2}}
  \end{overpic}
\end{minipage}  
\hspace{1mm}
\begin{minipage}{0.235\textwidth}
  \begin{overpic}
        [width = .95\textwidth]{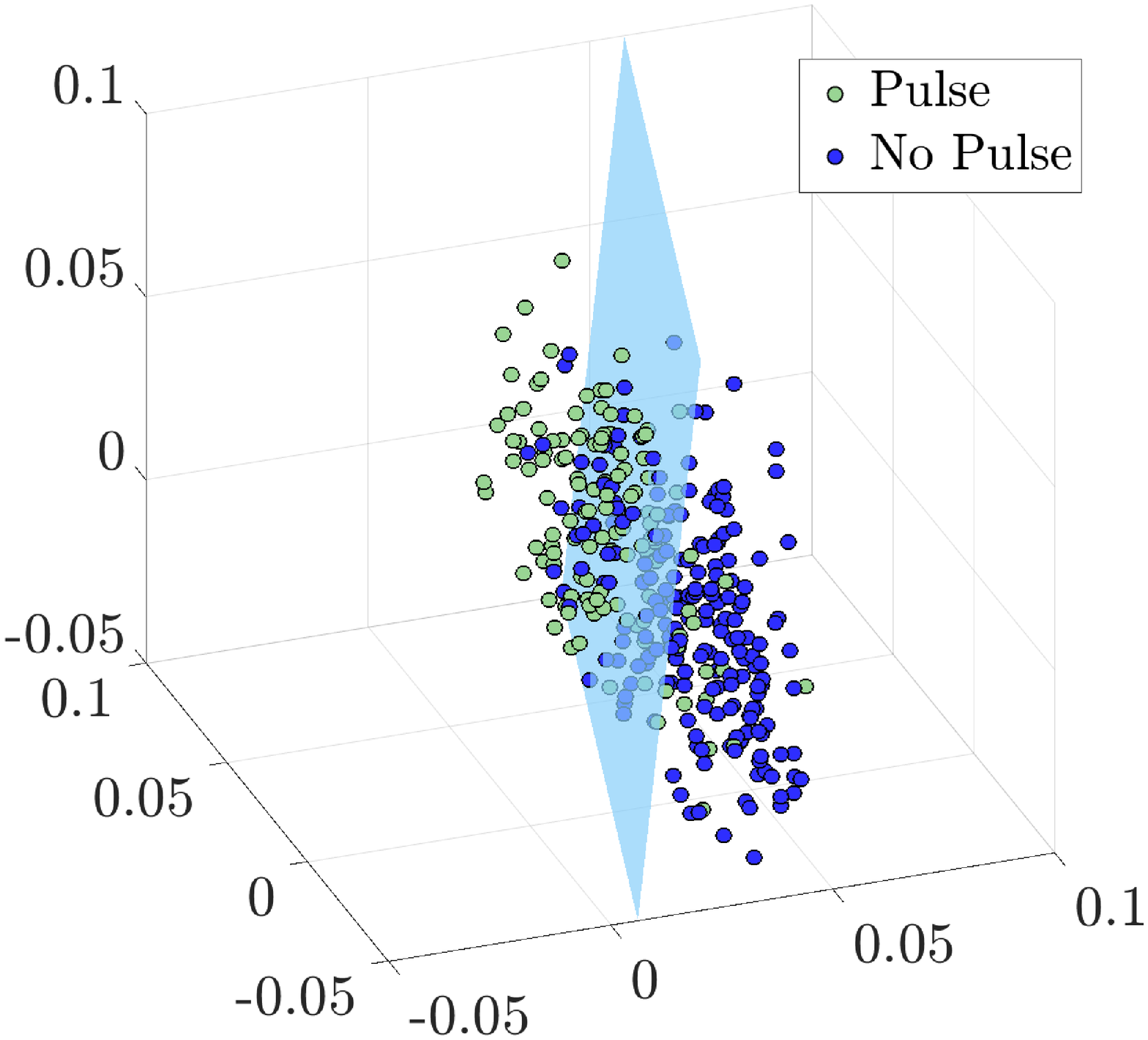}
        \put(-6,55){\rotatebox{90}{\small Mode 3}}
        \put(60,0){\rotatebox{7}{\small Mode 1}}
        \put(-4,30){\rotatebox{-55}{\small Mode 2}}
  \end{overpic} 
\end{minipage}
\caption{Left: Labeled training set with hyperplane. Right: Labeled test set with the same hyperplane.}
\label{fig:hyperplane}
\end{figure}

We used these three modes derived from the PCA analysis and heart rate as candidate features to train classifiers. In projecting the PCA modes onto a three dimensional space, we observe two distinct clusters: one with \textit{Pulse} labels, the other with \textit{No Pulse} labels. Thus, we can view these three modes as dominant features of the system. The formation of clusters in a low-dimensional feature space suggests amenability to discriminant analysis for which a number of classification models would be suitable \cite{bishop2006pattern,brunton2019data}. These models fall under the broader aegis of supervised machine learning and are critical in automating data-driven discovery processes. Therefore, we compared receiver operator curves with 5-fold cross validation using linear discriminant analysis (LDA), quadratic discriminant analysis (QDA), and support vector machine (SVM) models. 
We observed that all classifiers had comparable performance in training (Table \ref{table:comp}). Therefore we selected the simplest model (LDA) as the final model to reduce the risk of overfitting, which specifically had a training performance of 79\% (Figure \ref{fig:hyperplane}) during CPR and 87\% without CPR.  Furthermore, we observed that the addition of heart rate as a feature did not improve AUC. Therefore, the final model incorporates PCA modes 1-3 as the sole features.  

\begin{table}[t]
\resizebox{\columnwidth}{!}{%
 \begin{tabular}{c |  c c| c c} 
    &\multicolumn{2}{c}{\textbf{CPR}} & \multicolumn{2}{c}{\textbf{No CPR}}\\
  Classifier& Modes 1-3 & Modes 1-3, HR &Modes 1-3 & Modes 1-3, HR\\
  \hline
\color{blue}{LDA}& \color{blue} {0.79 (0.76,0.83)}&\color{blue} {0.80 (0.76,0.84)}& \color{blue}{0.87 (0.84,0.90)}& \color{blue}{0.87 (0.84,0.90)}\\
QDA& 0.80 (0.77,0.84)& 0.81 (0.78,0.85)& 0.87 (0.84,0.93)& 0.88 (0.85,0.91)\\
SVM& 0.79 (0.75,0.83)& 0.77 (0.73,0.81) & 0.87 (0.84,0.90)& 0.87 (0.84,0.90)\\
GMM&  0.77 (0.73,0.81)& 0.67 (0.63,0.72) & 0.85 (0.82,0.88)&0.86 (0.83,0.89)\\
NN & 0.80 (0.76,0.83)&0.83 (0.79,0.86) & 0.87 (0.84,0.89)&0.89 (0.86,0.91)\\
ConvNN & 0.78 (0.75,0.81)& -- &0.77 (0.74,0.81) & --\\
\end{tabular}
}

\caption{Comparison of Training AUC values and 95\% confidence intervals for various models, where LDA : Linear Discriminant Analysis, QDA :Quadratic Discriminant Analysis, SVM: Support Vector Machine, GMM: Gaussian Mixture Model, NN: Neural Network, ConvNN: Convolutional Neural Network.}
\label{table:comp}
\end{table}

\subsection{Assessment of Classification Performance} 
After developing the algorithm using the training set, we assessed performance of the pulse prediction algorithm using the test set of ECG segments with and without CPR separately. Performance was characterized by area under the receiver operating curve (AUC).  The AUC values and 95\% confidence intervals were calculated using bootstrapping \cite{qin2008comparison}. 

\begin{figure*}[t]
\begin{minipage}{.5\textwidth}
   \begin{minipage}{1\textwidth}
  \centering
  \includegraphics[width = .8\textwidth]{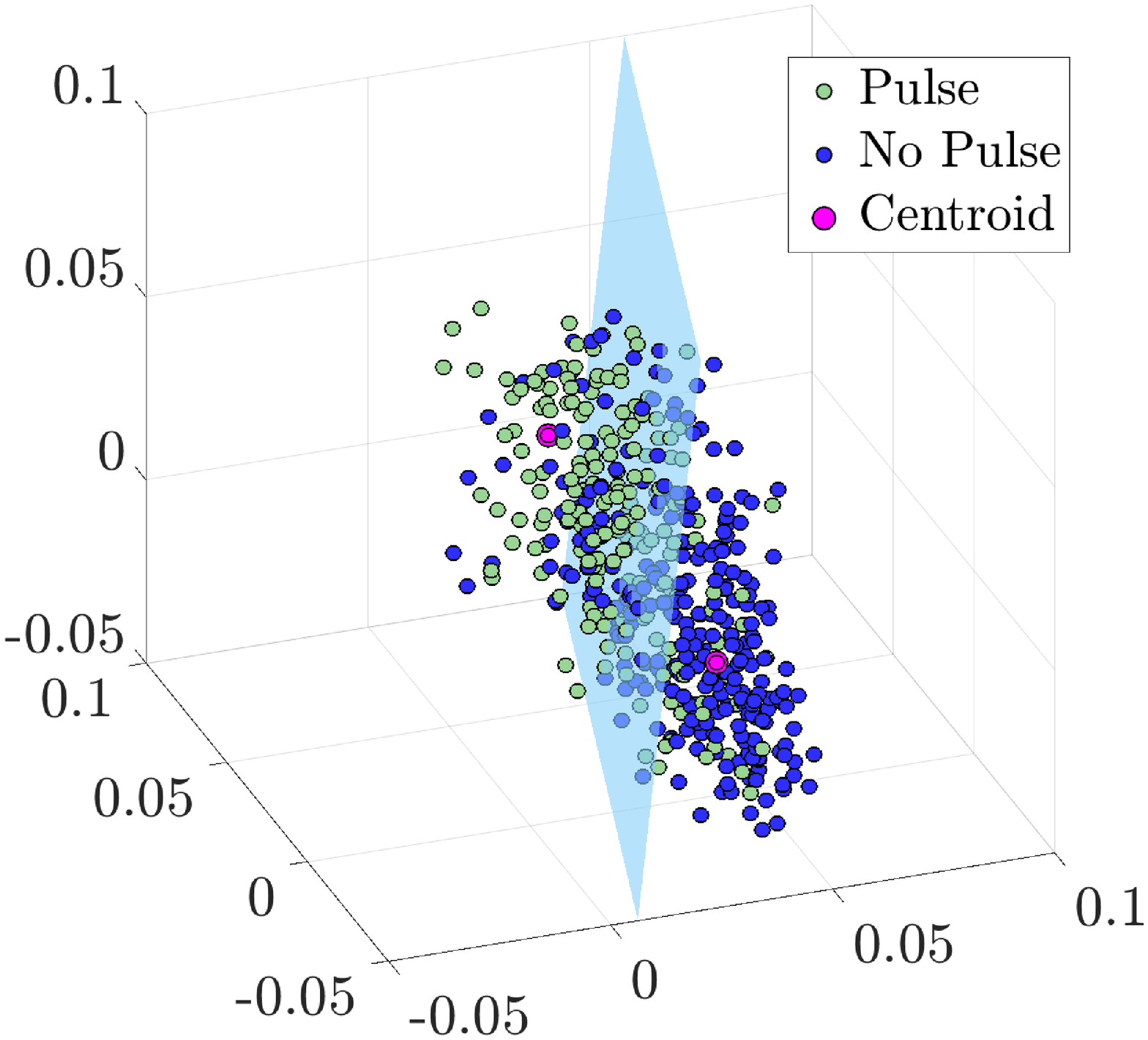}
  \linethickness{1.5pt}
  \put(-110,114){\color{Magenta}\vector(-3,-6){55}}
  \put(-80,75){\color{Magenta}\vector(3,-5){40}}
  \put(-80,10){\rotatebox{5}{Mode 1}}
  \put(-200,50){\rotatebox{-55}{Mode 2}}
  \put(-215,120){\rotatebox{90}{Mode 3}}
  \vspace{.5cm}
  \end{minipage}
     \begin{minipage}{.48\textwidth}
     \centering
     \begin{overpic}
    [width = 1 \textwidth]{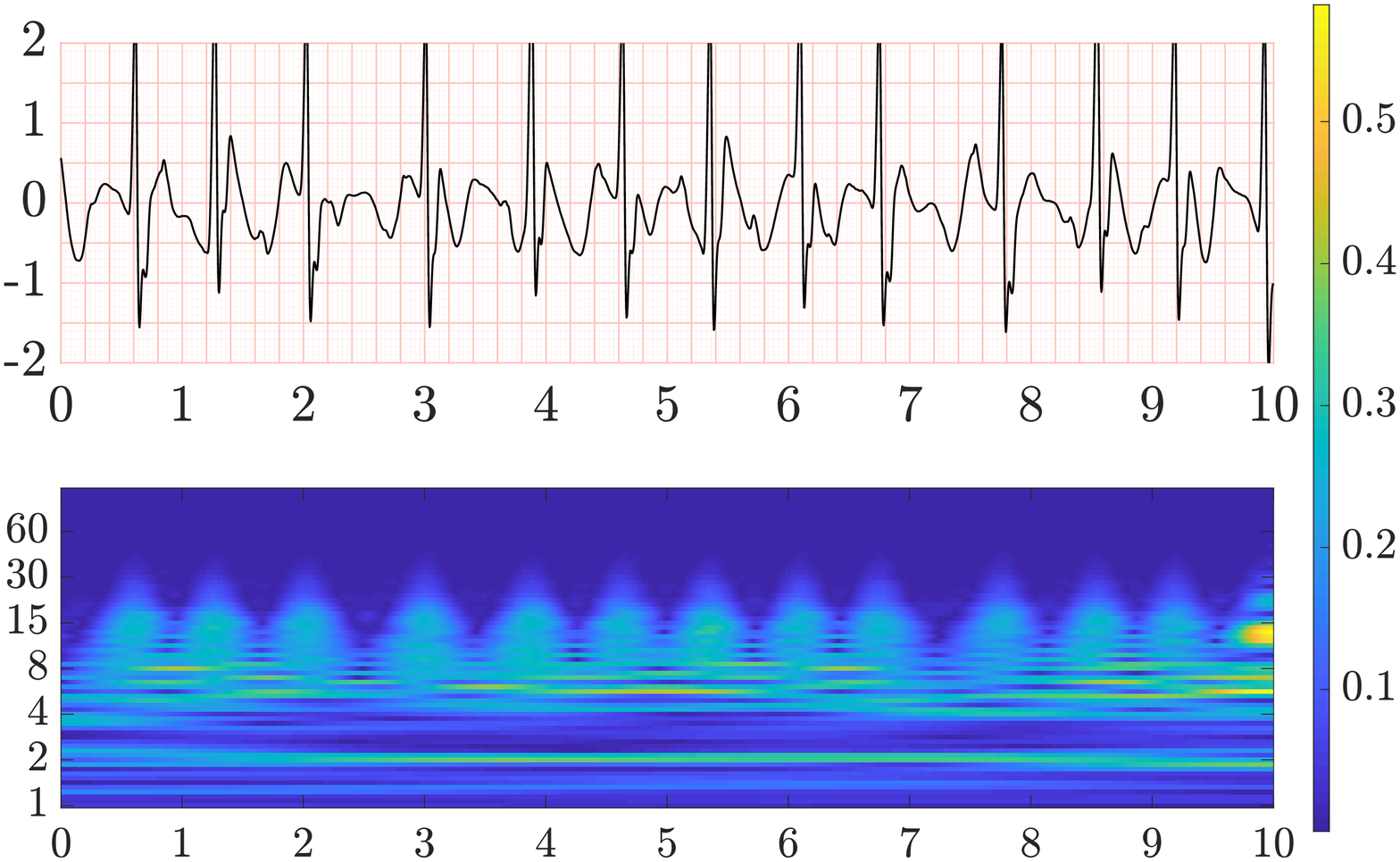}
    \put(30,68){\color{ForestGreen}{Representative}}
    \put(40,60){\color{ForestGreen}{Pulse}}
    \put(-5,7){\rotatebox{90}{\scriptsize \textit f (Hz)}}
    \put(-5,32){\rotatebox{90}{\scriptsize Voltage(mV)}}
    \put(40,-5){Time(s)}
    \end{overpic}
    \end{minipage}
    \hspace{1mm}
    \begin{minipage}{.48\textwidth}
    \centering
    \begin{overpic}
    [width = 1\textwidth]{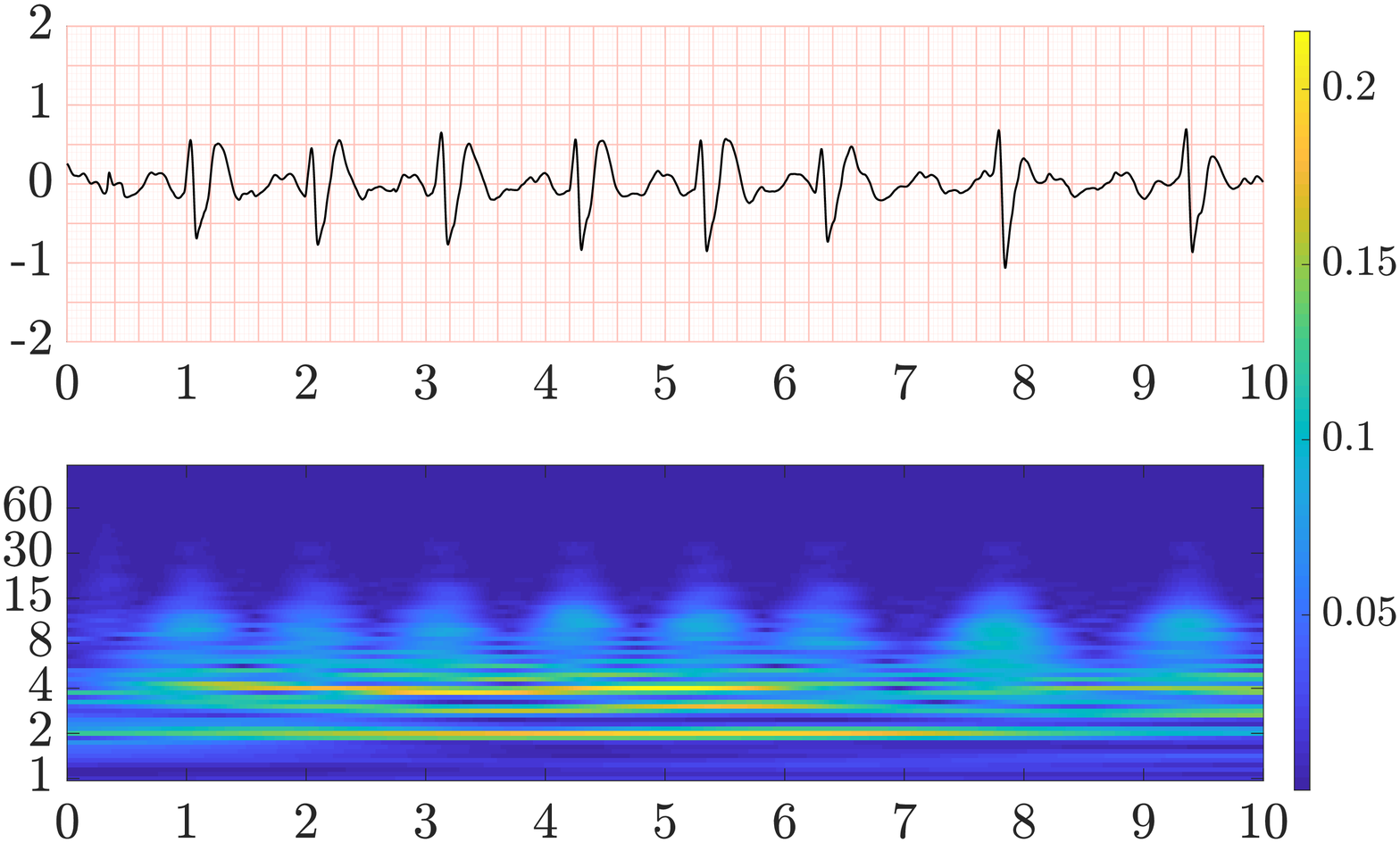}
    \put(35,68){\color{Blue}{Representative}}
    \put(40,60){\color{Blue}{No Pulse}}
    \put(40,-5){Time(s)}
    \end{overpic}
    \end{minipage}
    \end{minipage}
    \hspace{1mm}
    \begin{minipage}{.5\textwidth}
   \begin{minipage}{1\textwidth}
  \centering
  \includegraphics[width = .8\textwidth]{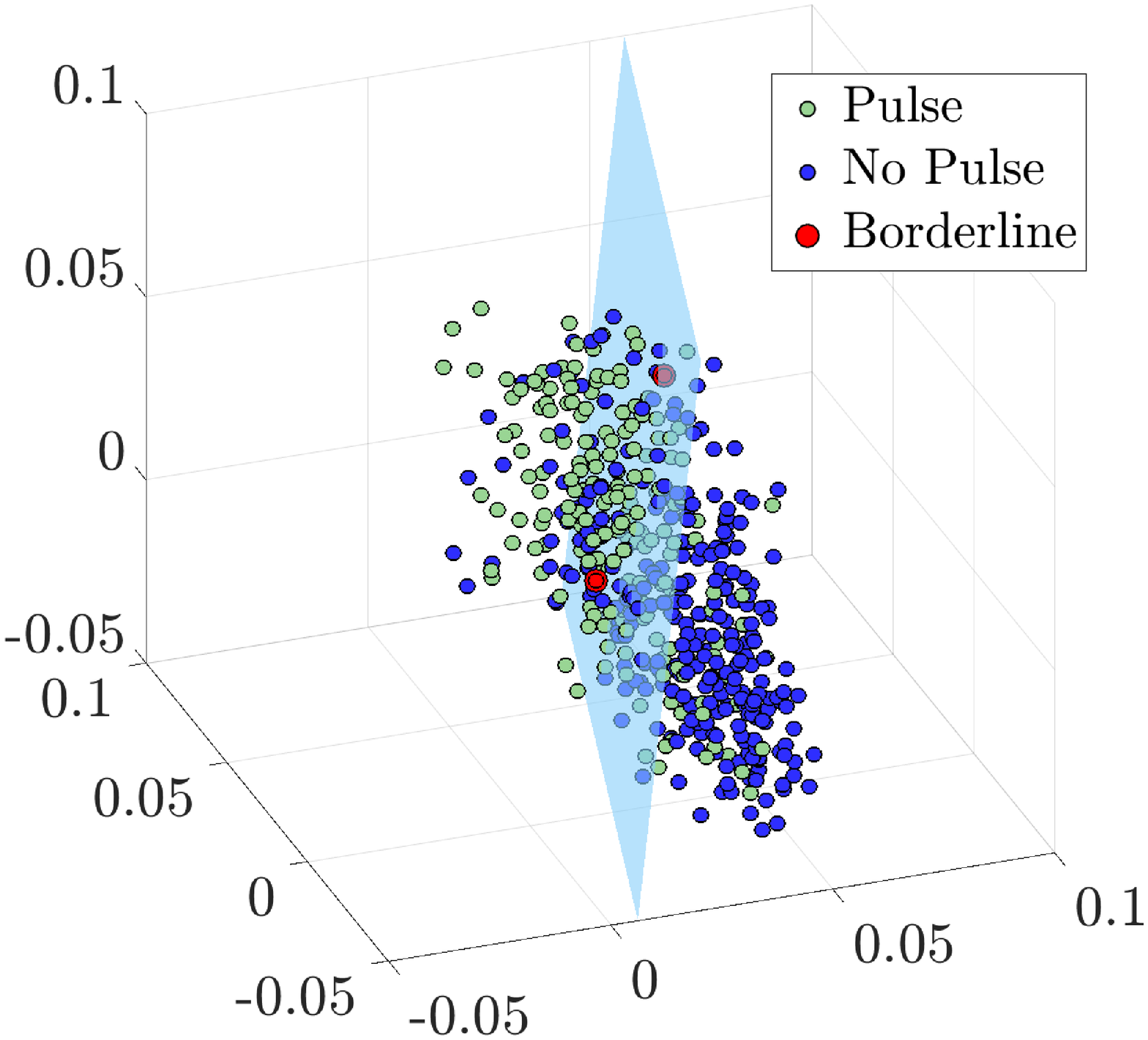}
  \linethickness{1.5pt}
  \put(-101,88){\color{red}\vector(-3,-4){65}}
  \put(-88,125){\color{red}\vector(3,-6){60}}
    \put(-80,10){\rotatebox{5}{Mode 1}}
  \put(-200,50){\rotatebox{-55}{Mode 2}}
  \put(-215,120){\rotatebox{90}{Mode 3}}
  \vspace{.5cm}
  \end{minipage}
     \begin{minipage}{.48\textwidth}
     \centering
     \begin{overpic}
    [width = 1 \textwidth]{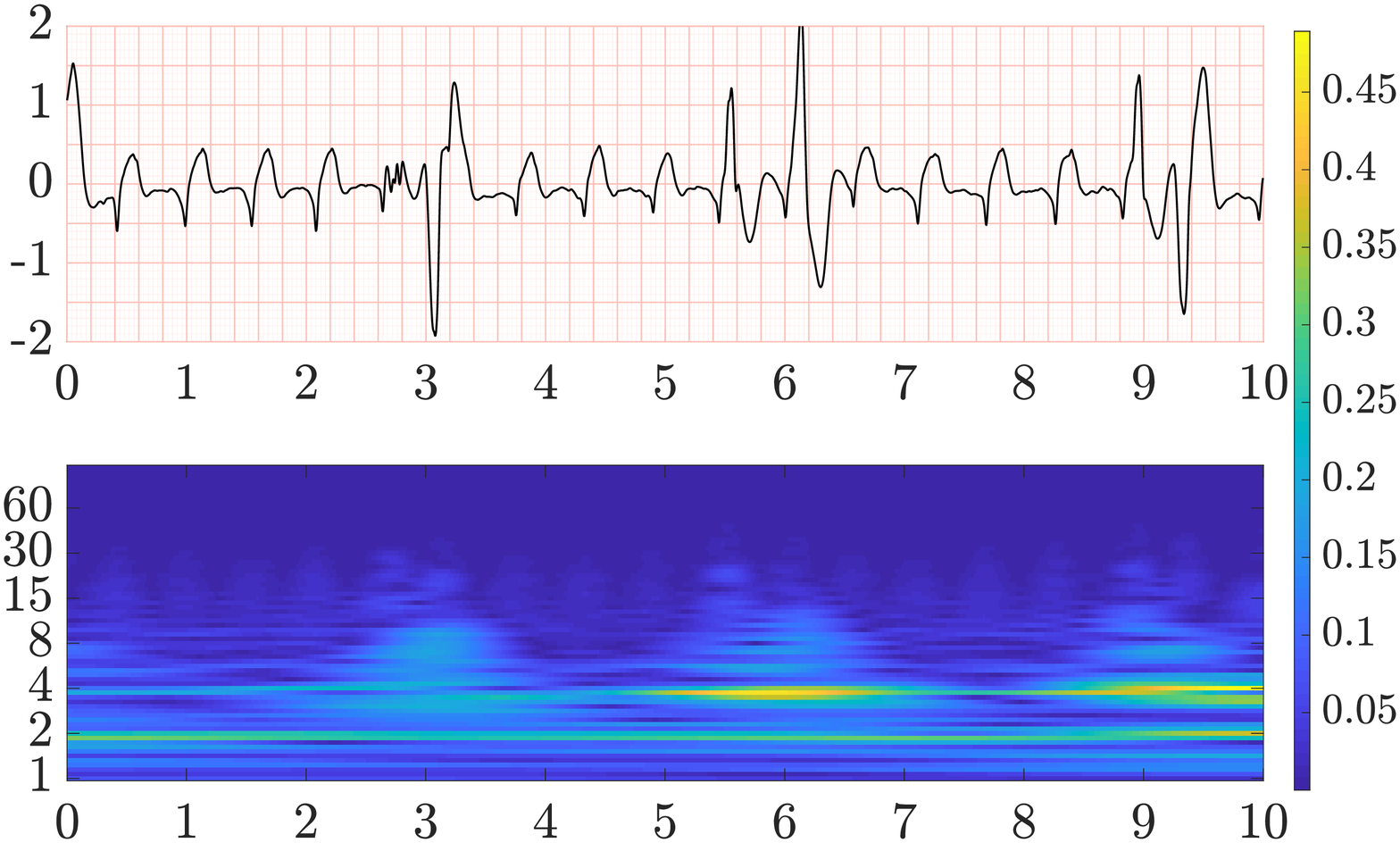}
    \put(35,68){\color{ForestGreen}{Borderline}}
    \put(40,60){\color{ForestGreen}{Pulse}}
    \put(40,-5){Time(s)}
    \end{overpic}
    \end{minipage}
    \hspace{1mm}
    \begin{minipage}{.48\textwidth}
    \centering
    \begin{overpic}
    [width = 1\textwidth]{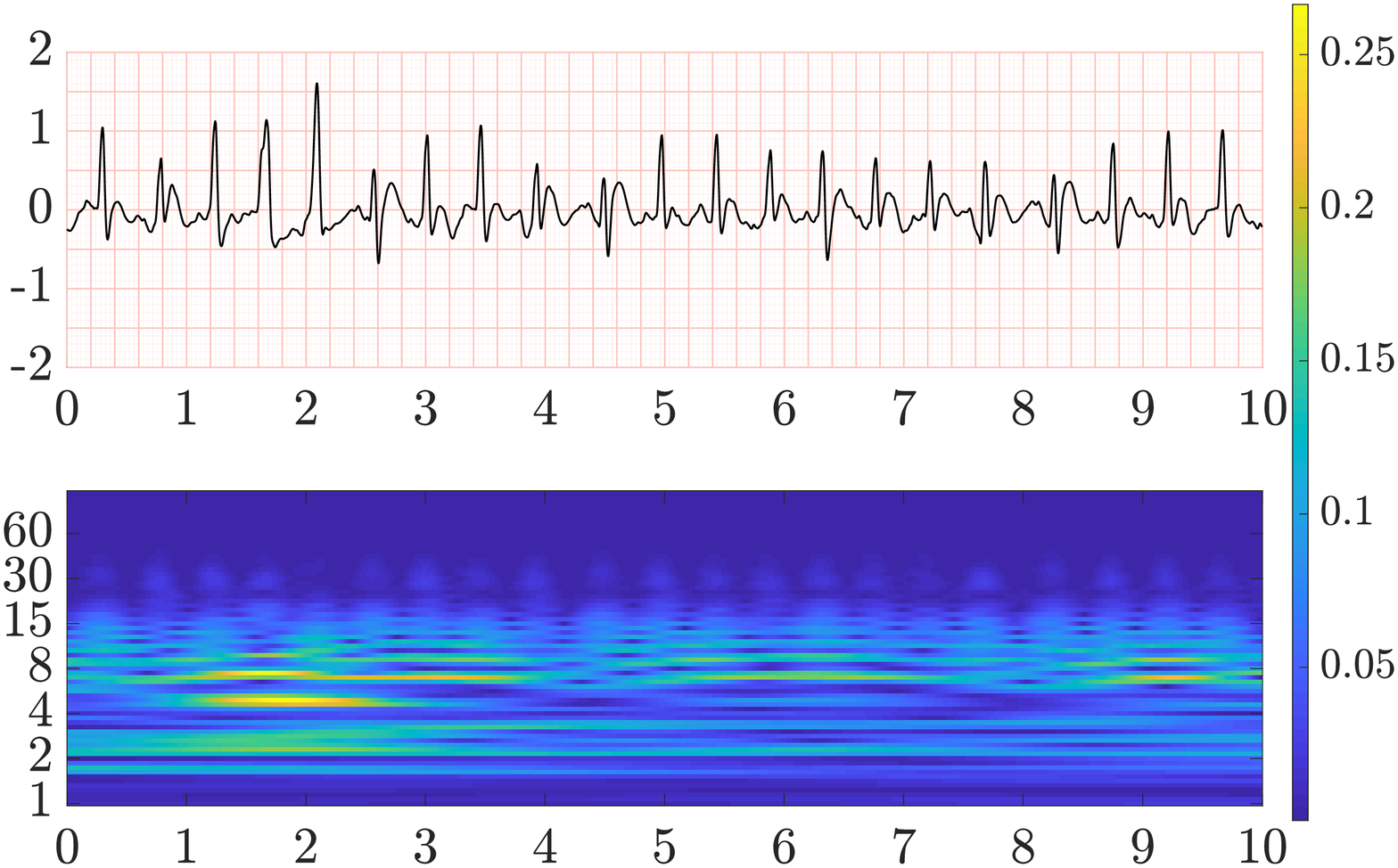}
    \put(40,68){\color{Blue}{Borderline}}
    \put(40,60){\color{Blue}{No Pulse}}
    \put(40,-5){Time(s)}
    \end{overpic}
    \end{minipage}
    \end{minipage}
    \vspace*{0.5cm}
    \caption{Left: Centroids of Cluster for Pulse and No Pulse labels respectively, Right:  Borderline points for Pulse and No Pulse respectively}
        \label{fig:centroids}

    \end{figure*}

\begin{table}[H]
\begin{center}
 \begin{tabular}{c | c c} 
Modes & CPR  & No CPR\\[0.5ex] 
 \hline
1-3 & 0.84 (0.797,0.88)&0.89 (.8550,.9189)\\ 
\\
\end{tabular}
\end{center}
\caption{Test AUC values for clips with CPR and no CPR respectively using Linear Discriminant Analysis (LDA)}
\label{table:test_noHR}
\end{table}
\section{Results}
We observed that the algorithm predicted the presence of pulse both with and without CPR. Specifically, AUC to discriminate between Pulse and Pulseless segments was 0.84 with CPR and 0.89 without CPR on test data (Table \ref{table:test_noHR}).



\section{Discussion}

An understanding of the ECG rhythm state and its physiological consequences provides the basis for using ECG to predict pulse.  The most common rhythms observed during cardiac arrest are ventricular fibrillation, asystole, and organized rhythms.  Ventricular fibrillation is characterized by chaotic, disorganized electrical activity on the ECG; and asystole by the absence of any electrical activity.  Neither produces a pulse and their presence on the ECG obviates the need for a pulse check.  Organized rhythms are defined by coordinated electrical activity on the ECG, but may or may not be associated with cardiac muscle contraction and a pulse (the latter commonly referred to as pulseless electrical activity).  Thus when observed on the ECG, organized rhythms require a pulse assessment as to the continued need for CPR and resuscitative efforts.         

 At present, automated tools to reliably identify a spontaneous pulse during CPR are not available \cite{ochoa1998competence}.Ideally, such a tool would leverage pre-existing monitoring technologies that are already in widespread use, such as the electrocardiogram (ECG), be independent of provider skill, and not interrupt CPR.  One challenge is that artifact from chest compressions can prevent accurate ECG analysis during chest compressions \cite{affatato2016see,ruiz2014rhythm}. Hence, prior methods to automatically detect pulse using the ECG have primarily been developed for use without CPR \cite{elola2019ecg,elola2019deep,losert2007thoracic,risdal2007automatic}. In recent work however, we developed an algorithm which predicts pulse status in real time from the ECG during CPR among patients with underlying organized rhythms \cite{kwok2020electrocardiogram}. This recent algorithm used a logistic combination of manually-designed features of  heart rate and energy within specific frequency bands.

 The method described in the current investigation integrates novel machine learning and signal processing techniques (see Fig \ref{fig:overview}) and presents a more scalable alternative to the recent prior method for pulse detection during CPR \cite{kwok2020electrocardiogram}. The earlier method relied upon three manually-designed features that were motivated by clinical experience: a measure of QRS rate and two measures of QRS amplitude and morphology (median magnitudes within two frequency bands centered at 6.25 and 25 Hz).  A limitation of that approach was that it did not provide a framework to model more subtle morphologic characteristics nor to discover new feature spaces.  In contrast, the components of the algorithm developed in the current method are modular and flexible; thus, new feature spaces and clustering techniques can be readily integrated and explored within the algorithmic structure proposed.  As larger quantities of data become available, data-based approaches such as deep neural networks could easily be integrated into the current method to further-improve performance.  Importantly, the current algorithm can be applied in a real-time manner during resuscitation, as the signal processing and feature extraction steps are not computationally intensive.  This minimization of computational energy could allow the algorithm to be incorporated into devices currently in production and enable real-time feedback to rescuers during CPR.  

 The algorithm described in the current study incorporates three spatial modes of the reduced wavelet scalogram, which enables some limited interpretability of the underlying variance of the two classes of data.  Examination of the spatial modes demonstrates a band in most of the modes around approximately 1.5-2 Hz (Fig. \ref{fig:singular_modes_two}).  This fundamental band and its harmonics may represent some effects of CPR (guidelines recommend CPR rates of 100-120 compression per minute)\cite{amann2010reduction} , but could also represent underlying heart rate. Heart rate may be intrinsically incorporated in these three modes, which could account for why the addition of heart rate as a separate feature did not appreciably improve model performance versus use of modes 1-3 alone. In modes 1 to 3, there is also a significant amount of energy at higher frequencies. However, in contrast, mode 4 exhibits minimal energy above the 2 Hz band other than the first and second harmonics (approximately 4 and 6 Hz).  Because each mode represents a linear combination of the high-dimensional scalogram, there is not a definitive relationship between the modes and conventionally-understood ECG characteristics (e.g, amplitude, QRS width, QRS rate, and rhythmicity).  However, further insight can be gained by visualizing representative cases of the temporal modes of this analysis. For example, a case that maps to the cluster centroid of the Pulse classification has QRS complexes that are more narrow and larger in amplitude than the cluster centroid of the No Pulse classification, as well as a faster rate (see Fig. \ref{fig:centroids}). Narrower QRS complexes could be indicative of increased myocardial conduction velocity, and thus may potentially be represented by increased broad-band high-frequency content such as that observed in modes 1-3. 

\subsection{Limitations}
In order to be deployed in clinical practice, the algorithm needs to be integrated with a function that can identify rhythms during CPR \cite{affatato2016see,ruiz2014rhythm,kwok2015adaptive}, because it assumes that the underlying rhythm is organized.  In addition, further study is necessary to determine how this algorithm would affect rescuer actions and resuscitation metrics such as CPR fraction and drug administration. Finally, while the study cohort included patients with initial rhythms of both ventricular fibrillation and pulseless electrical activity, the data were obtained from a single EMS system with generally good outcomes, so it is possible that algorithm performance may differ in other systems and populations.  

\subsection{Future Directions}
There are many ongoing challenges and promising directions that motivate future research in this area.  The raw ECG data-streams remain noisy and have frequent artifacts.  Machine learning architectures can potentially be integrated at the front end of methods in order to help provide more robust filtering of the ECG time-series data.  Further, as recording technologies improve and more data is available, more powerful neural network architectures can be used in the classification step to drive accuracies even higher.  Although the wavelet decomposition proved to be highly advantageous for extracting key features of the pulse status, there is a potential for improving the feature space used for extraction of dominant spatio-temporal signals responsible for accurate clustering and thus improving model performance.  Regardless, the data-driven discovery pipeline proposed here provides a baseline model for which improvements can be readily integrated and executed.

\section{Conclusion}
In summary, we developed a machine learning framework for automating the accurate prediction of pulse status from the ECG time-series.  The proposed algorithmic architecture integrates and leverages three mathematical architectures: (i) time-series analysis through a wavelet decomposition, (ii) feature engineering through dominant correlated spatio-temporal patterns, and (iii) clustering and classification methods.  We demonstrated a nearly 0.85 AUC value in identifying pulse status even while CPR is being performed on a patient, which has potential to provide essential feedback to rescuers during uninterrupted CPR.  The mathematical approach provides a data-driven framework that can be applied widely across current technology, giving a generalizable approach for clinical pulse prediction to potentially improve outcomes following cardiac arrest.   
Indeed, this approach suggests an opportunity to improve resuscitation through real-time identification of pulse status to minimize CPR interruptions and inform treatment decisions.


\section{Acknowledgments}
We appreciate the emergency medical services personnel of King County, without whom this study would not be possible. Informative discussions on pulse detection were conducted with Dawn Jorgensen, Stacy Gehman, and Chenguang Liu (Philips Healthcare, Bothell, WA). This study was supported in part by grants provided to the University of Washington by the American Heart Association (DS, HK, TR, PK) and the Washington Research Foundation (JC).

\section{Code Availability}
The following algorithm was implemented on MATLAB (R2018b). The code can be accessed using the following link: https://github.com/dsashid/PulsePrediction.

\bibliographystyle{IEEEtran}
\bibliography{Sashidhar_paper_PulsePredict.bib}

\end{document}